\documentclass[10pt]{article}
\usepackage{opex3,amsfonts,amsmath,amssymb,bm}
\usepackage{graphicx,color,subfigure}

\def\XXint#1#2#3{{\setbox0=\hbox{$#1{#2#3}{\int}$} 
     \vcenter{\hbox{$#2#3$}}\kern-.5\wd0}}

\def\beq{\begin{equation}}
\def\eeq{\end{equation}}
\def\bfg{\begin{figure}}
\def\efg{\end{figure}}

\def\address #1{ \begin{center}\small\it #1 \end{center} }

\begin{document}

\title{Graphical retrieval method for orthorhombic anisotropic materials}
\author{Simin Feng${}^1$}
\address{${}^1$Research and Intelligence Department, Physics Branch,  \\  Naval Air Warfare Center, China Lake, CA 93555}
\email{simin.feng@navy.mil}

\begin{abstract*}
We apply the equivalent theory to orthorhombic anisotropic materials and provide a general unit-cell design criterion for achieving a length-independent retrieval of the effective material parameters from a single layer of unit cells.  We introduce a graphical retrieval method and phase unwrapping techniques.  The graphical method utilizes the linear regression technique.  Our method can reduce the uncertainty of experimental measurements and the ambiguity of phase unwrapping.  Moreover, the graphical method can simultaneously determine the bulk values of the six effective material parameters, permittivity and permeability tensors, from a single layer of unit cells.  \\
\end{abstract*}




\section{Introduction}
Metamaterials (MMs) are artificial materials engineered to achieve unusual electromagnetic (EM) properties that are not normally found in nature\cite{Fang}-\cite{Feng}.  A comprehensive review can be found in the book\cite{Capolino}.  Distinguished from photonic crystals, metamaterials are made from periodical structures with unit cells much smaller than the wavelength of light.  A general method to construct large area 3D MMs is layer-by-layer fabrication technique\cite{Valentine2}-\cite{Shalaev}.  In stacked MMs, interaction between adjacent layers makes it difficult to extract bulk material parameters from a single unit-cell layer.  It has been found that the retrieved effective metamaterial parameters are often dependent on the number of unit cells along the propagation direction\cite{Valentine2, Andryieuski}.  It appears that there is no clear methodology on how to accurately predict the bulk values of the effective permittivities and permeabilies through a single layer of unit cells.  Currently there is no simple and effective way to resolve phase ambiguity in determining the phase of the transmission and reflection of electromagnetic fields because of phase wrapping.  Although the phase ambiguity is a common issue for the parameter retrieval of the general composite materials, this issue becomes more significant for metamaterials where typically resonances, positive refractive index, and negative refractive index are all present in the same frequency band.  In this paper, we apply Herpin's equivalent theorem\cite{Herpin} to orthorhombic anisotropic media and provide a simple way to accurately predict the effective bulk material parameters from a single layer of unit cells.  We introduce a graphical retrieval method and phase unwrapping techniques, which can simultaneously determine the six material parameters, the permittivity and permeability tensors, from one unit cell.

\section{Equivalent theory}
Layer-by-layer fabrication method renders metamaterials intrinsically anisotropic.  In MMs one unit-cell layer is usually composed of several sub-layers of different materials or nanostructures.  We limit our discussion in nonchiral materials.  To illustrate the key point we adopt multilayer method that has been used to model MMs by many groups\cite{Smith}-\cite{Chen}.  We further assume that the principal axes of the sub-layers are parallel in each direction.  In the principal coordinate system, the permittivity and permeability tensors are given by
\begin{equation}
\label{epmu}
\bar{\bar\epsilon}_n = \begin{pmatrix} \epsilon_{nx} & 0 & 0 \\ 0 & \epsilon_{ny} & 0 \\ 0 & 0 & \epsilon_{nz} \end{pmatrix}  \,, \hspace{.3in}
\bar{\bar\mu}_n = \begin{pmatrix} \mu_{nx} & 0 & 0 \\ 0 & \mu_{ny} & 0 \\ 0 & 0 & \mu_{nz} \end{pmatrix}  \,,
\end{equation}
where $n=1,2,\cdots$.  The scalar terms $\epsilon_{nj}$ and $\mu_{nj}\ (j=x,y,z)$ are complex.  Consider a monochromatic wave of frequency $\omega$ with time dependence $\exp(-i\omega t)$ propagates inside the orthorhombic anisotropic materials.  In each layer we have
\begin{eqnarray}
\label{Maxw}
\begin{split}
\nabla\times\bigl( \bar{\bar\mu}_n^{-1} \cdot \nabla\times{\bm E}\bigr)  &=\, k_0^2\bigl(\bar{\bar\epsilon}_n \cdot{\bm E}\bigr)  \,,\\
\nabla\times\bigl( \bar{\bar\epsilon}_n^{-1} \cdot \nabla\times{\bm H}\bigr)  &=\, k_0^2\bigl(\bar{\bar\mu}_n \cdot{\bm H}\bigr)  \,,
\end{split}
\end{eqnarray}
where $k_0=\omega/c$.  If the plane of incidence is one of the crystal planes, the TE and TM polarizations are decoupled.

\begin{figure}[htb]
\centering\includegraphics[width=.33\textwidth]{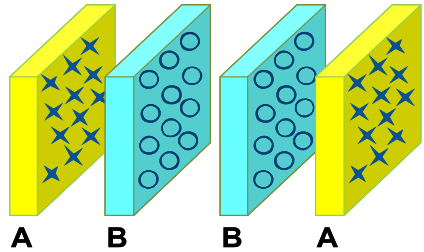}
\caption{A schematic shows how to construct a symmetric unit cell, which is composed of two original asymmetric unit cells.}
\label{UnitCell}
\end{figure}

\noindent
According to Herpin's equivalent theorem\cite{Herpin}, every general multilayer is equivalent to a two-homogeneous-layer system and every symmetric multilayer is equivalent to a single homogeneous layer, characterized by an equivalent index and equivalent thickness\cite{Epstein}.  In metamaterials, a single unit-cell layer can be considered as a multilayer system.  So it is equivalent to a two-homogeneous-layer system, denoted as $AB$.  We can then construct a symmetric unit cell by cascading two unit cells as $ABBA$, which is equivalent to a single homogeneous layer according to Herpin's theorem.  This process is illustrated in Fig.~\ref{UnitCell}.  Applying this methodology, a symmetric unit-cell layer can often be constructed regardless the number of sub-layers and complexity of each sub-layer in the original unit cell.  The permittivities and permeabilities retrieved from the symmetric unit cell, which is composed of two original asymmetric unit cells, will represent the bulk material parameters.  Thus, a length-independent description can be achieved.  \\

To make it more clear, let $\bm{M}$ represents the characteristic matrix of one symmetric unit cell.  According to Herpin's equivalent theorem, it can be replaced by an equivalent single layer.  Assume the x-z plane is the plane of incidence.  For TM mode, i.e. $\bm{H}=(0,H_y,0)$ and $\bm{E}=(E_x,0,E_z)$, we have
\begin{equation}
\label{Herpin}
\begin{split}
M_{11} &= M_{22} = \cos\psi_e  \,, \\
M_{12} &= \frac{\sin\psi_e}{i{\cal Z}_e} \,, \hskip.1in  M_{21} = -i{\cal Z}_e \sin\psi_e  \,, \\
\end{split}
\end{equation}
where $\psi_e$ is the equivalent phase thickness of the symmetric unit cell;  ${\cal Z}_e$ is the equivalent impedance.  If the material contains $N$-layer symmetric unit cells, the characteristic matrix of the material is given by
\begin{equation}
\label{MatxN}
\begin{pmatrix} \cos\psi_e  &  \dfrac{1}{i{\cal Z}_e}\sin\psi_e  \\  -i{\cal Z}_e\sin\psi_e  &  \cos\psi_e  \end{pmatrix}^N =
\begin{pmatrix} \cos(N\psi_e)  &  \dfrac{1}{i{\cal Z}_e}\sin(N\psi_e)  \\  -i{\cal Z}_e\sin(N\psi_e)  &  \cos(N\psi_e)  \end{pmatrix} \,.
\end{equation}
A similar expression for TE mode can be obtained by replacing the ${\cal Z}_e$ with the negative admittance $-{\cal Y}_e$.  Here the ${\cal Z}_e$ and ${\cal Y}_e$ are, respectively, the generalized impedance and admittance because they include incidence angle, i.e., Eq.~(\ref{MatxN}) is valid for both normal and oblique incidence.  They are given by
\begin{equation}
\label{Disp}
\begin{split}
&\mbox{TM:}\hspace{.3in} 	{\cal Z}_e = \frac{k_z}{k_0\epsilon_x}  \,,\hspace{.3in}
	k_z^2 = k_0^2\epsilon_x\,\mu_y - \frac{\epsilon_x}{\epsilon_z} k_x^2   \,, \\
&\mbox{TE:}\hspace{.3in} 	{\cal Y}_e = \frac{k_z}{k_0\mu_x}  \,,\hspace{.3in}
	k_z^2 = k_0^2\epsilon_y\,\mu_x - \frac{\mu_x}{\mu_z} k_x^2  \,, \\
\end{split}
\end{equation}
where $(\epsilon_x,\epsilon_y,\epsilon_z)$ and $(\mu_x,\mu_y,\mu_z)$ are, respectively, the effective permittivities and permeabilities of the equivalent layer.  As shown in Eq.~(\ref{MatxN}), the total phase of a $N$-layer system equals $N$ times the phase $\psi_e$ of the single layer;  whereas the impedance ${\cal Z}_e$ or admittance ${\cal Y}_e$ is independent of the number of layers.  These properties imply that the material parameters retrieved from the $N$ layers of unit cells are the same as those retrieved from one layer of unit cells.  In other words, the bulk metamaterial parameters can be predicted from a single symmetric unit cell.  Note that the characteristic matrix of a N-layer asymmetric unit cells does not have the nice properties as those in Eq.~(\ref{MatxN}).  As a consequence, the retrieved material parameters will be dependent on the number of unit cells along the propagation direction.  Hence, the bulk material parameters cannot be predicted from one unit-cell layer.  In other words, a length-independent description cannot be achieved for asymmetric unit cells.   \\

\section{Graphical retrieval method}
In this section, we will introduce a graphical retrieval method based on Herpin's theorem and the previous method\cite{Smith}.  From above discussion a symmetric unit cell can be represented by three equivalent layers $ABA$.  Let the $z$-direction perpendicular to the plane of the layers.  Assume each layer is orthorhombic anisotropic with the principal axes parallel in each direction and the effective permittivity and permeability tensors are given by Eq.~(\ref{epmu}).  Let the x-z plane be the plane of incidence.  The characteristic matrix of TM mode is defined as
\begin{equation}
\begin{pmatrix} H^i_y \vspace{1mm} \\ E^i_x \end{pmatrix} = {\bm M} \begin{pmatrix} H^o_y \vspace{1mm} \\ E^o_x \end{pmatrix}  =
\begin{pmatrix} M_{11} & M_{12} \vspace{1mm} \\  M_{21} & M_{22} \end{pmatrix} \begin{pmatrix} H^o_y \vspace{1mm} \\ E^o_x \end{pmatrix}  \,,
\end{equation}
where `i' refers to the input; and `o' refers to the output.  By matching boundary condition, the relationship between the scattering and characteristic matrices of a unit cell is given by
\begin{equation}
\label{MinS}
\begin{split}
M_{11} &= M_{22} = \frac{\Bigl[\bigl(1+S_{21}\bigr)\bigl(1-S_{12}\bigr) + S_{11}S_{22}\Bigr]} {2S_{11}}  \,,\\
M_{12} &= \frac{\Bigl[\bigl(1+S_{21}\bigr)\bigl(1+S_{12}\bigr) - S_{11}S_{22}\Bigr]} {2S_{11}{\cal Z}_o}  \,,\\
M_{21} &= \frac{{\cal Z}_i}{2S_{11}} \Bigl[\bigl(1-S_{21}\bigr)\bigl(1-S_{12}\bigr) - S_{11}S_{22}\Bigr]  \,.
\end{split}
\end{equation}
where ${\cal Z}_i$ and ${\cal Z}_o$ are, respectively, the generalized input and output impedance whose values depend on the incidence angle and background permittivity and permeability at the input and output sides.  In our definition, $S_{ij} (i=j)$ is related to the transmission and $S_{ij} (i\ne j)$ is related to the reflection.  That means, for the TE polarization $S_{11}$ and $S_{21}$ are, respectively, the transmission and reflection coefficients of the electric field; whereas for the TM polarization the transmission and reflection coefficients of the electric field are given by $\dfrac{\epsilon_1k_{2z}}{\epsilon_2k_{1z}}S_{11}$ and $-S_{21}$, respectively.  Note that as long as the layer has inversion symmetry, $M_{11}=M_{22}$ regardless the input and output impedances.  For scattering matrix, $S_{11}=S_{22}$ only if ${\cal Z}_i={\cal Z}_o$.  In addition if $M_{11}=M_{22}$, we then also have $S_{12}=S_{21}$.  From Eq.~(\ref{Herpin}), the dispersion relation and the effective impedance ${\cal Z}_e$ of the unit cell $ABA$ are given by
\begin{equation}
\label{DispZe}
\begin{split}
& \cos(K_ed)  =  M_{11} = \cos(k_{1z}d_1)\cos(k_{2z}d_2) - \eta^+ \sin(k_{1z}d_1)\sin(k_{2z}d_2)  \,,\\
& {\cal Z}_e^2 = \frac{M_{21}}{M_{12}} = {\cal Z}_1^2 
		\frac{\sin(k_{1z}d_1)\cos(k_{2z}d_2) + \eta^+\cos(k_{1z}d_1)\sin(k_{2z}d_2) - \eta^-\sin(k_{2z}d_2)}
		{\sin(k_{1z}d_1)\cos(k_{2z}d_2) + \eta^+\cos(k_{1z}d_1)\sin(k_{2z}d_2) + \eta^-\sin(k_{2z}d_2)}  \,,
\end{split}
\end{equation}
where the phase $\psi_e=K_ed$ and the $K_e$ is the effective propagation constant.  The subscript 1 and 2 refer to the layer A and B, respectively.  The ${\cal Z}_1$ is the generalized impedance of the equivalent layer A.  The z-component wave vector $k_{1z}$ and $k_{2z}$ are determined from the dispersion relations in Eq.~(\ref{Disp}).  The $d_1$ is twice the thickness of layer A;  and the $d_2$ is the thickness of layer B.  The period of the symmetric unit cell is $d=d_1+d_2$.  Other parameters in Eq.~(\ref{DispZe}) are given by
\begin{equation}
\label{eta}
\begin{split}
\eta^\pm  &= \frac{1}{2} \left(\frac{\epsilon_{2x}\,k_{1z}} {\epsilon_{1x}\,k_{2z}} \pm \frac{\epsilon_{1x}\,k_{2z}} {\epsilon_{2x}\,k_{1z}}\right) 
	\hspace{.25in}\mbox{for TM} \,, \\
\eta^\pm  &= \frac{1}{2} \left(\frac{\mu_{2x}\,k_{1z}} {\mu_{1x}\,k_{2z}} \pm \frac{\mu_{1x}\,k_{2z}} {\mu_{2x}\,k_{1z}}\right) 
	\hspace{.25in}\mbox{for TE}  \,.
\end{split}
\end{equation}
Although some MMs are mesoscopic media, nevertheless we restrict our consideration to those satisfying $d\ll\lambda$, after tedious derivation, Eq.~(\ref{DispZe}) can be simplified to
\begin{equation}
\label{TM-TE}
\begin{split}
&\mbox{TM:}\hspace{.3in} \frac{K_e^2}{k_0^2} = \overline\epsilon_x\,\overline\mu_y - \frac{\overline\epsilon_x}{\overline\epsilon_z} \frac{k_x^2}{k_0^2}
	\,,\hspace{.3in}  {\cal Z}_e^2 = \frac{\overline\mu_y}{\overline\epsilon_x} - \frac{k_x^2}{k_0^2\overline\epsilon_x\,\overline\epsilon_z}  \,,\\
&\mbox{TE:}\hspace{.3in} \frac{K_e^2}{k_0^2} = \overline\epsilon_y\,\overline\mu_x - \frac{\overline\mu_x}{\overline\mu_z} \frac{k_x^2}{k_0^2}
	\,,\hspace{.3in}  {\cal Y}_e^2 = \frac{\overline\epsilon_y}{\overline\mu_x} - \frac{k_x^2}{k_0^2\overline\mu_x\,\overline\mu_z}  \,.
\end{split}
\end{equation}
Here, the $\overline\epsilon_j$ and $\overline\mu_j\ (j=x,y,z)$ are the bulk values of the effective permittivities and permeabilities, respectively.  They are related to the material parameters of the equivalent layers A and B through
\begin{equation}
\label{epmuEff}
\begin{split}
\overline\epsilon_p &= \frac{d_1\epsilon_{1p} + d_2\epsilon_{2p}}{d}  \,,  \hspace{.4in}
\overline\epsilon_z = \frac{d\,\epsilon_{1z}\,\epsilon_{2z}}{d_1\epsilon_{2z} + d_2\epsilon_{1z}}  \,, \\
\overline\mu_p &= \frac{d_1\mu_{1p} + d_2\mu_{2p}}{d}  \,,  \hspace{.4in}
\overline\mu_z = \frac{d\,\mu_{1z}\,\mu_{2z}} {d_1\mu_{2z} + d_2\mu_{1z}}  \,,
\end{split}
\end{equation}
where $p=x,y$.  Equation~(\ref{epmuEff}) is a simplification of the most general linear case treated by Lakhtakia\cite{Lakhtakia}.  In practice, the material parameters of the equivalent layers are unknown.  One can only measure scattering parameters.  In most experiments ${\cal Z}_i={\cal Z}_o$, combining Eqs.~(\ref{MinS}) and~(\ref{TM-TE}), we obtained the retrieval formula:
\begin{equation}
\label{RetvLine}
\begin{split}
&\mbox{TM:}\hskip.35in  Y_M = \overline\epsilon_x\,\overline\mu_y - \frac{\overline\epsilon_x}{\overline\epsilon_z}X  \,,\hskip.3in
		Y_m = \frac{\overline\mu_y}{\overline\epsilon_x} - \frac{X}{\overline\epsilon_x\,\overline\epsilon_z}   \,, \\
&\mbox{TE:}\hskip.35in  Y_E = \overline\epsilon_y\,\overline\mu_x - \frac{\overline\mu_x}{\overline\mu_z}X  \,,\hskip.3in
		Y_e = \frac{\overline\epsilon_y}{\overline\mu_x} - \frac{X}{\overline\mu_x\,\overline\mu_z}  \,,
\end{split}
\end{equation}
where
\begin{eqnarray}
\label{X}  &&X \equiv \epsilon_b\mu_b\sin^2\theta  \,,\hspace{.2in}  Y_m\equiv \frac{\mu_b}{\epsilon_b} S\cos^2\theta  \,,\hspace{.2in}
					 Y_e\equiv \frac{\epsilon_b}{\mu_b} S\cos^2\theta  \,,\hspace{.2in}
					 S\equiv \frac{\bigl(1-S_{21}\bigr)^2 - S_{11}^2} {\bigl(1+S_{21}\bigr)^2 - S_{11}^2}  \,,\\
\label{Y}  &&Y_M = Y_E\equiv \left\{\frac{1}{k_0d} \left[2m\pi\pm \cos^{-1}\left(\frac{1-S_{21}^2+S_{11}^2}{2S_{11}}\right)\right]\right\}^2 
					 \,,\hspace{.2in}  m = 0,\pm1,\pm2,\cdots  \,,
\end{eqnarray}
where $\theta$ is the incidence angle.  The $\epsilon_b$ and $\mu_b$ are, respectively, the background relative permittivity and permeability.  The retrieval formulas in Eq.~(\ref{RetvLine}) provide four straight lines ($Y_M$, $Y_m$, $Y_E$, $Y_e$ vs. $X$), two for each polarization.  The $Y_M$ and $Y_E$ represent the dispersion lines for TM and TE polarizations, respectively.  The $Y_m$ and $Y_e$ are the corresponding impedance and admittance lines.  These straight lines are easy to implement experimentally.  After measuring the scattering parameters at several incidence angles and plotting the data according to Eqs.~(\ref{RetvLine})-(\ref{Y}), use linear regression technique to calculate the slopes and Y-intercepts of the four lines.  From the slopes and Y-intercepts, the six effective material parameters, $\overline\epsilon_j$ and $\overline\mu_j\ (j=x,y,z)$, can be retrieved simultaneously.  Let $Y_M^0$ and $Y_m^0$ represent the Y-intercepts of the two lines in TM polarization, $S_M$ and $S_m$ the corresponding slopes of the lines;  whereas $Y_E^0$, $Y_e^0$, $S_E$, and $S_e$ are the corresponding quantities for TE polarization.  Thus, 
\begin{equation}
\label{RetVal}
\begin{split}
\mbox{TM:} \hskip.2in &n_m = \pm\sqrt{Y_M^0} \,,\hskip.2in Z = \pm\sqrt{Y_m^0} \,,\hskip.2in \overline\epsilon_x = \frac{n_m}{Z}
	\,,\hskip.2in \overline\mu_y = n_mZ  \,,\hskip.2in  \overline\epsilon_z = -\frac{\overline\epsilon_x}{S_M}  \,,\\
\mbox{TE:} \hskip.2in &n_e = \pm\sqrt{Y_E^0} \,,\hskip.2in  Y = \pm\sqrt{Y_e^0} \,,\hskip.2in \overline\epsilon_y = n_eY
	\,,\hskip.2in \overline\mu_x = \frac{n_e}{Y}  \,,\hskip.2in  \overline\mu_z = -\frac{\overline\mu_x}{S_E}  \,,
\end{split}
\end{equation}
where the $\pm$ sign in Eq.~(\ref{RetVal}) can be fixed by requiring the imaginary part of refractive index greater than zero, i.e.
$\Im(n_m)>0$, $\Im(n_e)>0$, and the real part of impedance and admittance greater than zero, i.e. $\Re(Z)>0$, and $\Re(Y)>0$ for passive media\cite{Smith}.

\section{Resolving phase branch}
\label{Phase}
There are two issues regarding the branch determination in Eq.~(\ref{Y}), the $\pm$ sign in front of the inverse cosine and the phase branch $m$.  The inverse cosine $\cos^{-1}(\cdot)$ represents the principal value: $0\le\cos^{-1}(\cdot)\le\pi$.  Knowing only the cosine value, the phase angle cannot be determined.  One need the information of sine, which can be obtained from the imaginary part of the cosine as following:
\begin{equation}
A \equiv \frac{1-S_{21}^2+S_{11}^2}{2S_{11}} = \cos\varphi = \cos(\varphi_r+i\varphi_i) 
	= \cos\varphi_r\cosh\varphi_i - i\sin\varphi_r\sinh\varphi_i  \,,
\end{equation}
where $\varphi_r$ and $\varphi_i$ are real.  For passive medium, $\varphi_i>0$, and thus $\sinh\phi_i>0$.  Therefore, the $\pm$ sign in front of the inverse cosine in Eq.~(\ref{Y}) can be resolved from the imaginary part of $A$:
\begin{equation}
\label{pmSign}
\varphi =  \left\{ \begin{matrix} \cos^{-1}A  & &  \mbox{if  } \Im(A) < 0  \\
	2\pi - \cos^{-1}A  & &  \mbox{if } \Im(A) > 0  \end{matrix}  \right.  \,.
\end{equation}

After determining the $\pm$ sign, next step is to resolve the correct phase branch $m$.  This part can be very confusing when negative refractive index might be involved.  Notice that among the four retrieval lines in Eq.~(\ref{RetvLine}), only the dispersion lines, $Y_M\sim X$ and $Y_E\sim X$, are functions of the phase branch $m$, the impedance and admittance lines are independent of $m$.  Employing this property, we provide three methods that might help to resolve the correct phase branch $m$.  \\

{\it Method-1, Using $\overline\epsilon_x^2$ and $\overline\mu_x^2$}:\ \ 
For the TM polarization in Eq.~(\ref{RetvLine}), the $\overline\epsilon_x^2$ can be obtained either from the intercepts as $\overline\epsilon_x^2=\dfrac{Y_M^0}{Y_m^0}$ or from the slopes as $\overline\epsilon_x^2=\dfrac{S_M}{S_m}$, and the two results should be close.  When the branch is wrong, the two results can be significantly different.  Similarly for the TE polarization, the $\overline\mu_x^2$ can be obtained from either $\overline\mu_x^2=\dfrac{Y_E^0}{Y_e^0}$ or $\overline\mu_x^2=\dfrac{S_E}{S_e}$.  Hence, the correct branch $m$ is the one that minimizes the absolute value of the differences, i.e. 
\begin{equation}
\label{Branch1}
\begin{split}
\mbox{TM:} \hspace{.3in}  &\min\left\{\left|\dfrac{Y_M^0(m)}{Y_m^0}-\dfrac{S_M(m)}{S_m}\right|:\ \ m=0,\pm1,\pm2,\cdots\right\}  \,, \\
\mbox{TE:} \hspace{.3in}  &\min\left\{\left|\dfrac{Y_E^0(m)}{Y_e^0}-\dfrac{S_E(m)}{S_e}\right|:\ \ m=0,\pm1,\pm2,\cdots\right\}  \,.
\end{split}
\end{equation}

{\it Method-2, Using $\dfrac{\overline\mu_y}{\overline\epsilon_z}$ and $\dfrac{\overline\epsilon_y}{\overline\mu_z}$}:\ \ 
Alternatively, we can use $\dfrac{\overline\mu_y}{\overline\epsilon_z}$ for TM and $\dfrac{\overline\epsilon_y}{\overline\mu_z}$ for TE as criteria to determine the correct phase branch, i.e.
\begin{equation}
\label{Branch2}
\begin{split}
\mbox{TM:} \hspace{.3in}  &\min\left\{\left|Y_M^0(m) S_m - Y_m^0 S_M(m)\right|:\ \ m=0,\pm1,\pm2,\cdots\right\}  \,, \\
\mbox{TE:} \hspace{.3in}  &\min\left\{\left|Y_E^0(m) S_e - Y_e^0 S_E(m)\right|:\ \ m=0,\pm1,\pm2,\cdots\right\}  \,.
\end{split}
\end{equation}

{\it Method-3, Using $\overline\epsilon_z\,\overline\mu_y$ and $\overline\epsilon_y\,\overline\mu_z$}:\ \ 
The third method is to use $\overline\epsilon_z\,\overline\mu_y$ for TM and $\overline\epsilon_y\,\overline\mu_z$ for TE as criteria to select the correct phase branch.  Then, the algorithm becomes
\begin{equation}
\label{Branch3}
\begin{split}
\mbox{TM:} \hspace{.3in}  &\min\left\{\left|\frac{Y_M^0(m)}{S_M(m)} - \frac{Y_m^0}{S_m}\right|:\ \ m=0,\pm1,\pm2,\cdots\right\}  \,, \\
\mbox{TE:} \hspace{.3in}  &\min\left\{\left|\frac{Y_E^0(m)}{S_E(m)} - \frac{Y_e^0}{S_e}\right|:\ \ m=0,\pm1,\pm2,\cdots\right\}  \,.
\end{split}
\end{equation}
In above three methods, the branch number $m$ predicted by the first two methods is always consistent for all the frequencies in the several examples we tested.  The third method predicts the same result as the first two for most of the frequencies, but sometimes it can be different by $\pm1$ at the edge of phase transition frequencies in resonant regimes.  Note that before applying Eqs.~(\ref{Branch1}), (\ref{Branch2}), or~(\ref{Branch3}) to resolve the correct phase branch, the $\pm$ sign in front of the inverse cosine in Eq.~(\ref{Y}) must be determined first.  From the several examples we tested, using above branch-resolving techniques, there is no need to recourse adjacent frequencies in determining the correct phase branch.  Recoursing adjacent frequencies can become confusing when both positive and negative refractive indices are present in the same frequency band.

\begin{figure}[hbt]
\centering
\subfigure[TM polarization: dispersion lines (left panels) and impedance lines (right panels).]
{
    \label{RetrvLine:a}
    \includegraphics[width=.45\textwidth]{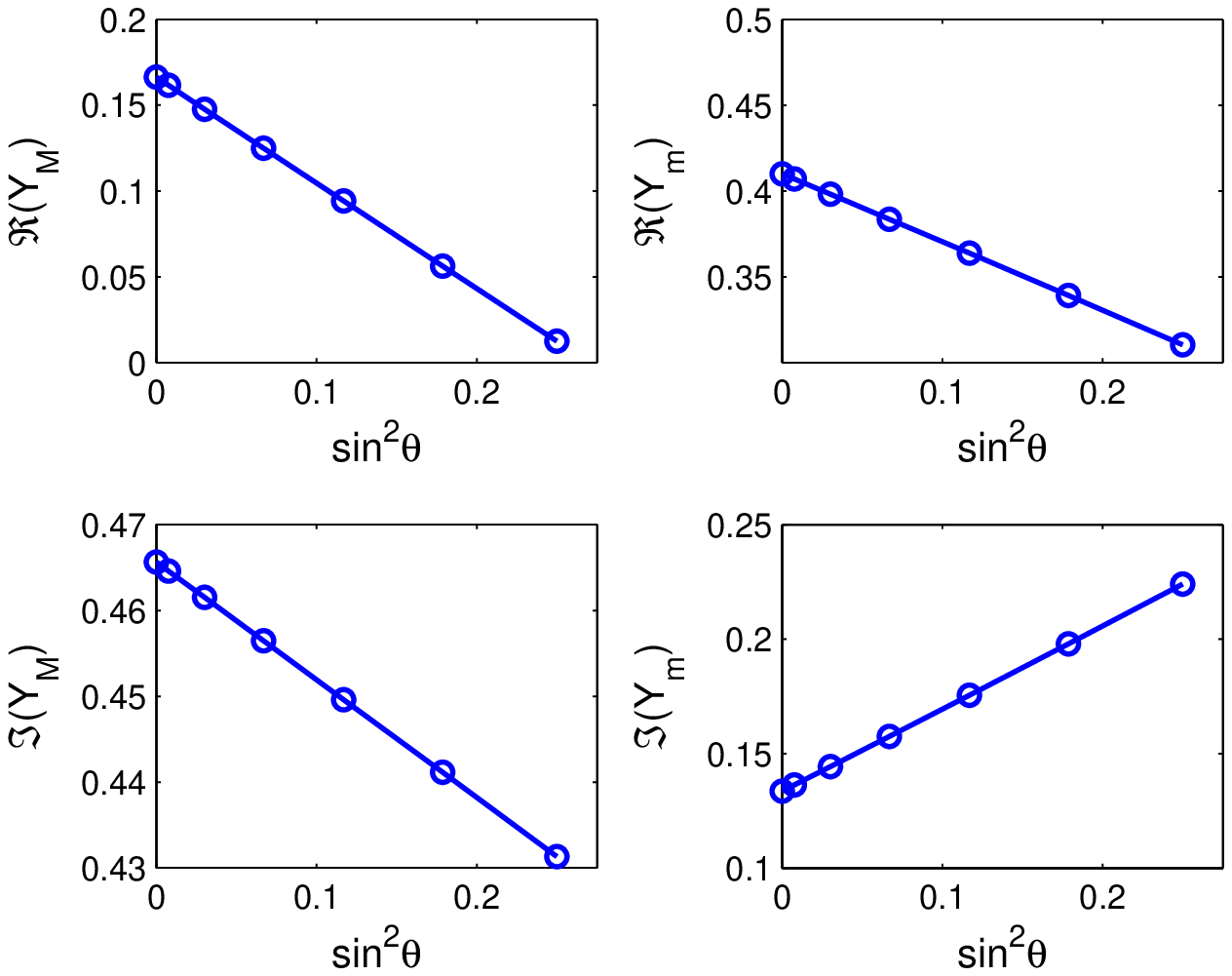}
}
\hspace{1cm}
\subfigure[TE polarization: dispersion lines (left panels) and admittance lines (right panels).]
{
    \label{RetrvLine:b}
    \includegraphics[width=.45\textwidth]{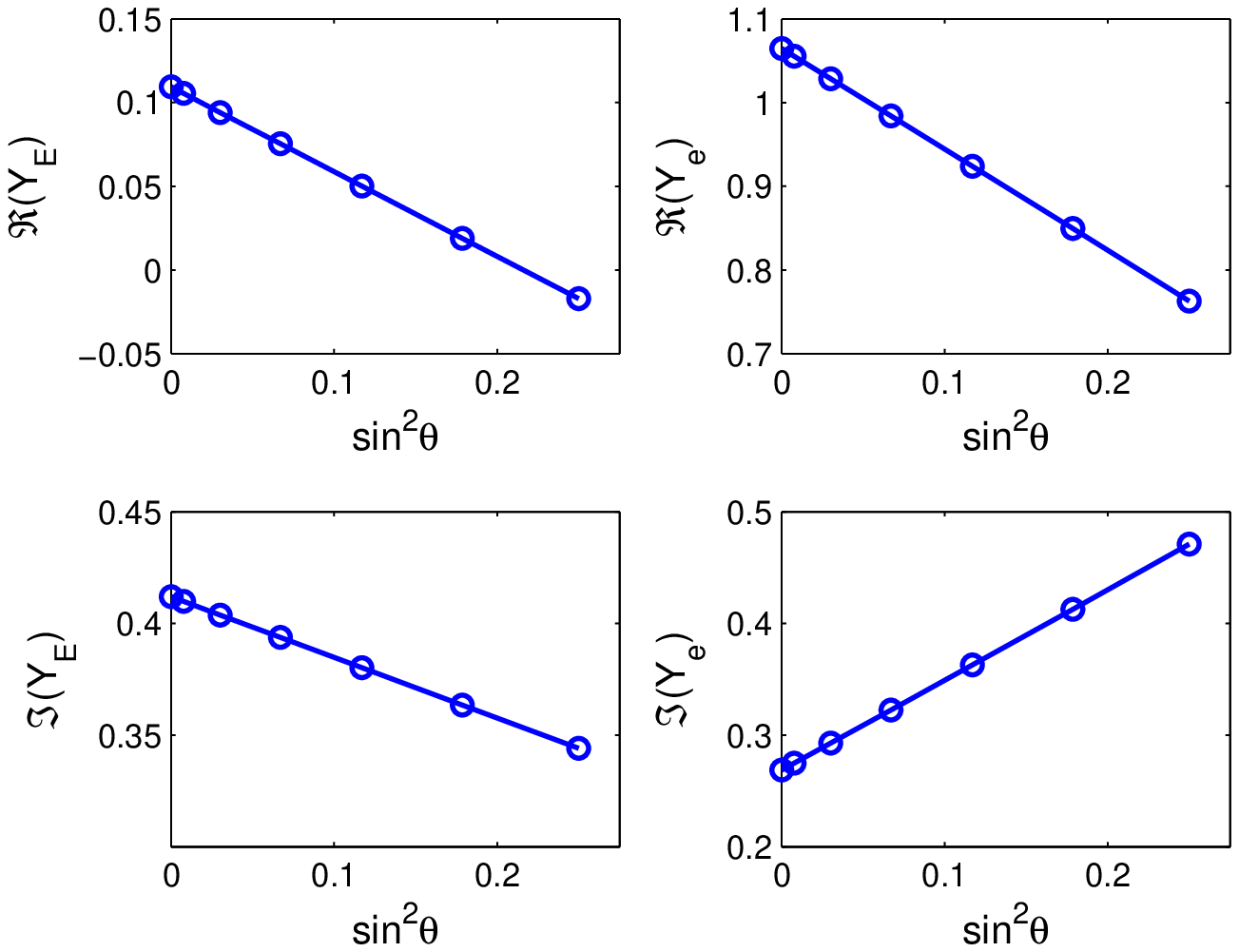}
}
\caption{Retrieval lines at the frequency $30\,$THz calculated from the scattering matrix of a single layer of unit cells at seven incidence angles (denoted as circles) uniformly distributed from $0$ to $30$ degree.  The horizontal axis is $\sin^2\theta$.  Background medium is vacuum.  Top panels: real part.  Bottom panels: imaginary part.}
\label{RetrvLine}
\end{figure}

\begin{figure}[hbt]
\centering
\subfigure[One-period thickness.]
{
	\label{Branch:a}
	\includegraphics[width=.45\textwidth]{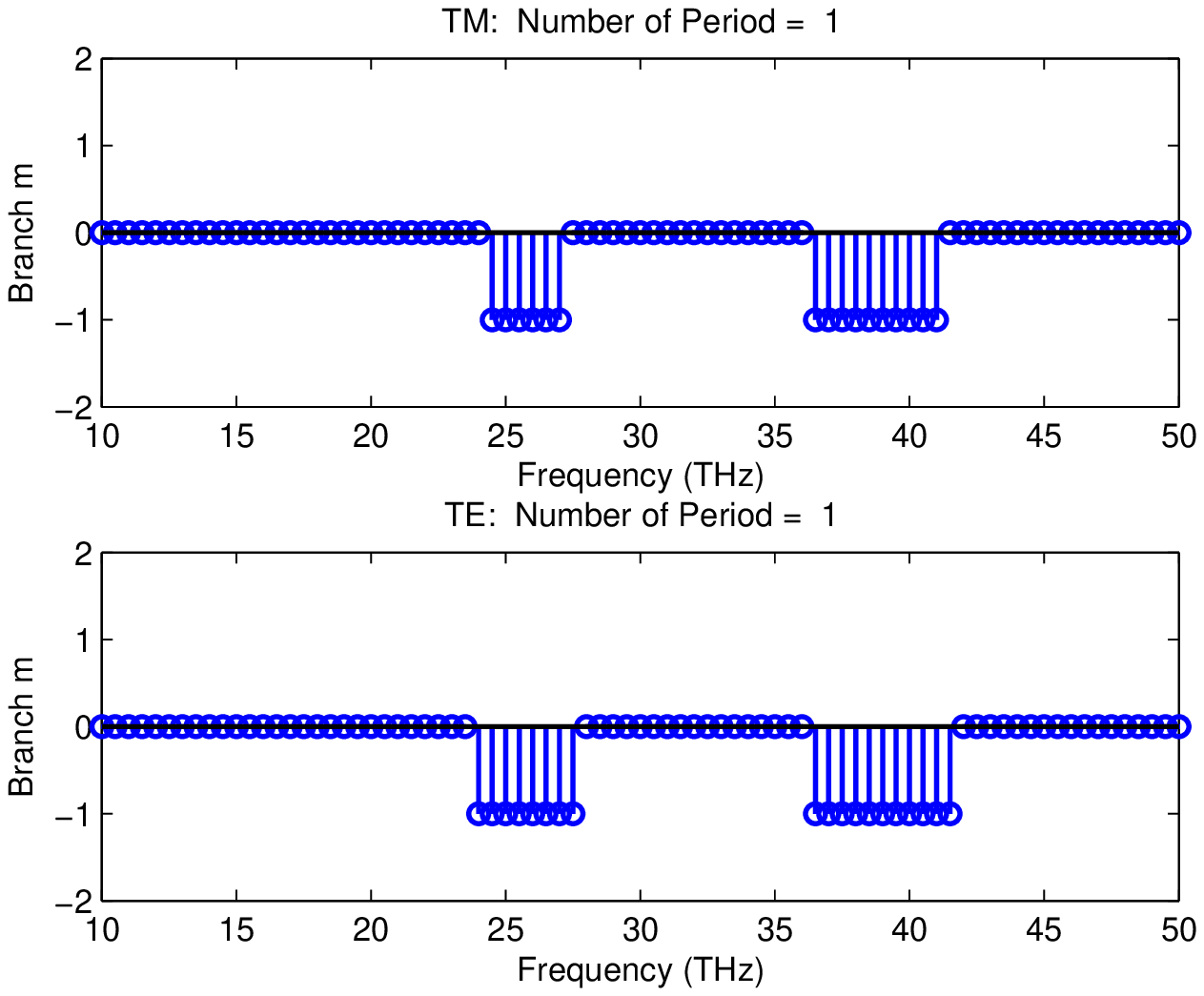}
}
\hspace{1cm}
\subfigure[Six-period thickness.]
{
	\label{Branch:b}
	\includegraphics[width=.45\textwidth]{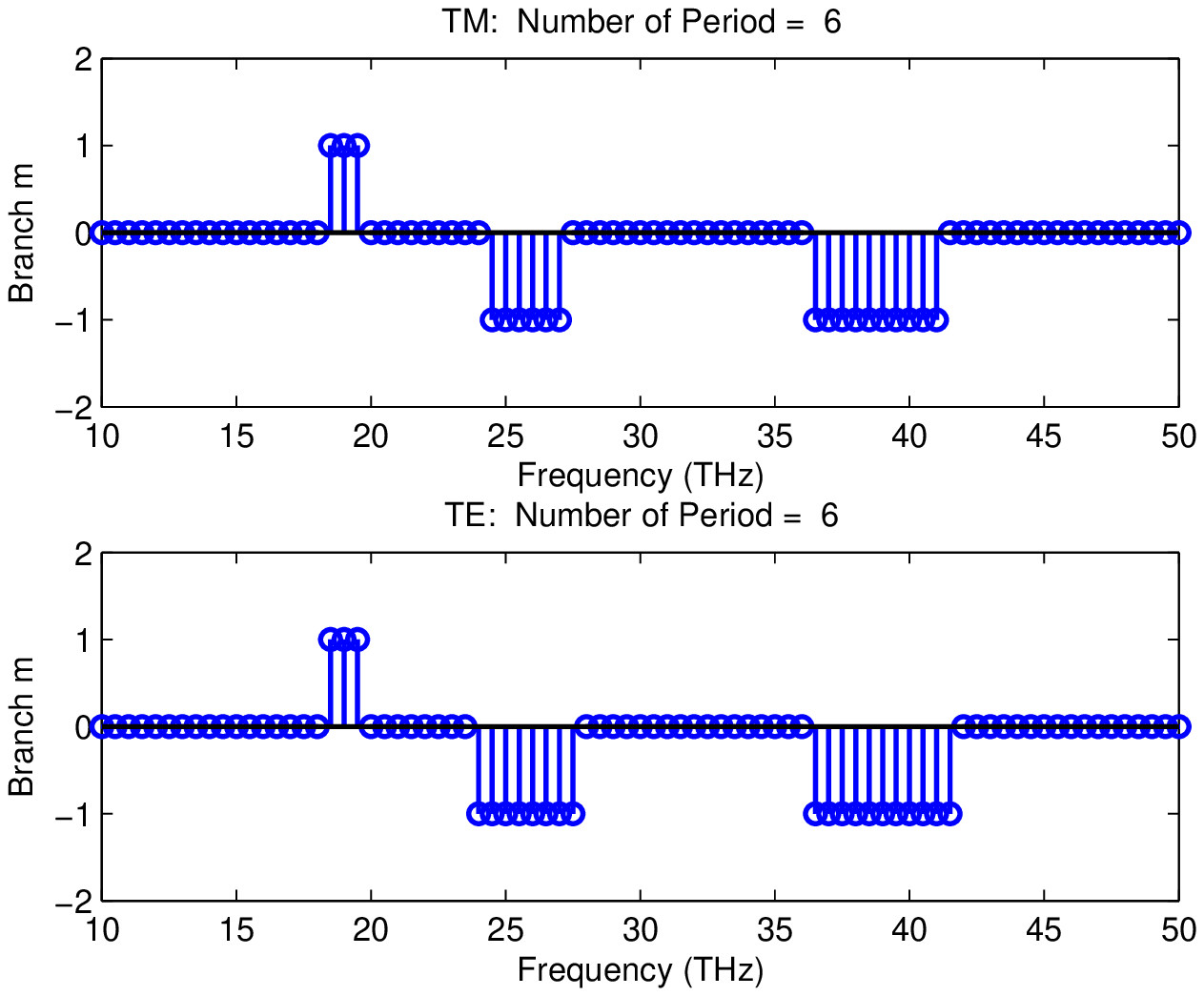}
}
\caption{Branch number predicted by the method-3 [Eq.~(\ref{Branch3})] vs. frequency.  Since $d\ll\lambda$, most of the frequencies are in the fundamental branch $m=0$.  The $m=-1$ indicates the frequencies of the negative refractive index [$n<0$, see Fig.~\ref{nZY:a}].  The $m=1$ corresponds to the frequencies of the high valves of the positive refractive index [see Fig.~\ref{nZY:a}].  Top: TM polarization.  Bottom: TE polarization.}
\label{Branch}
\end{figure}

\section{Discussion}
\begin{figure}[hbt]
\centering
\subfigure[Retrieved refractive index for TM polarization (left panels) and for TE polarization (right panels).]
{
	\label{nZY:a}
	\includegraphics[width=.45\textwidth]{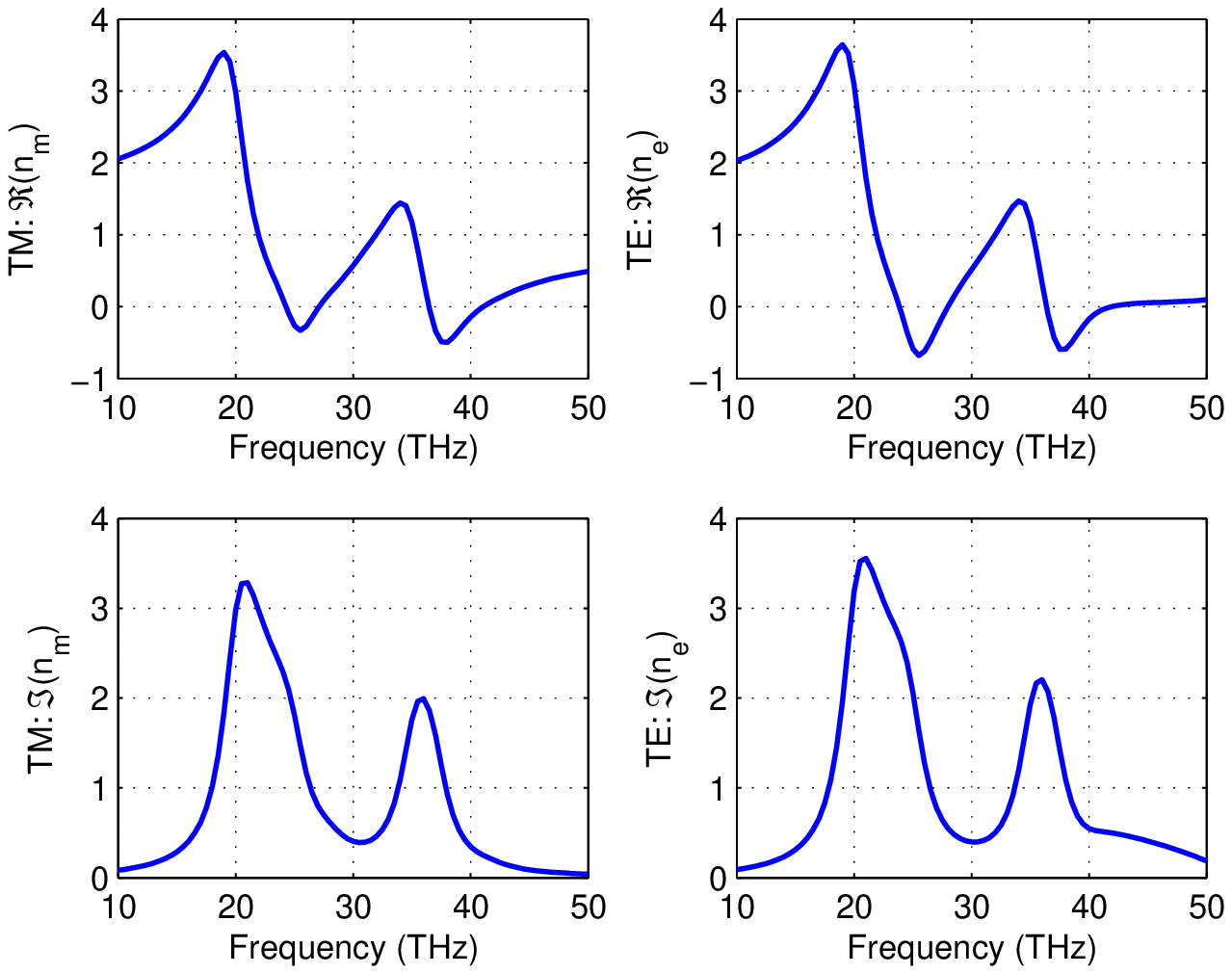}
}
\hspace{1cm}
\subfigure[Left panels: Retrieved impedance for TM polarization.  Right panels: Retrieved admittance for TE polarization.]
{
	\label{nZY:b}
	\includegraphics[width=.45\textwidth]{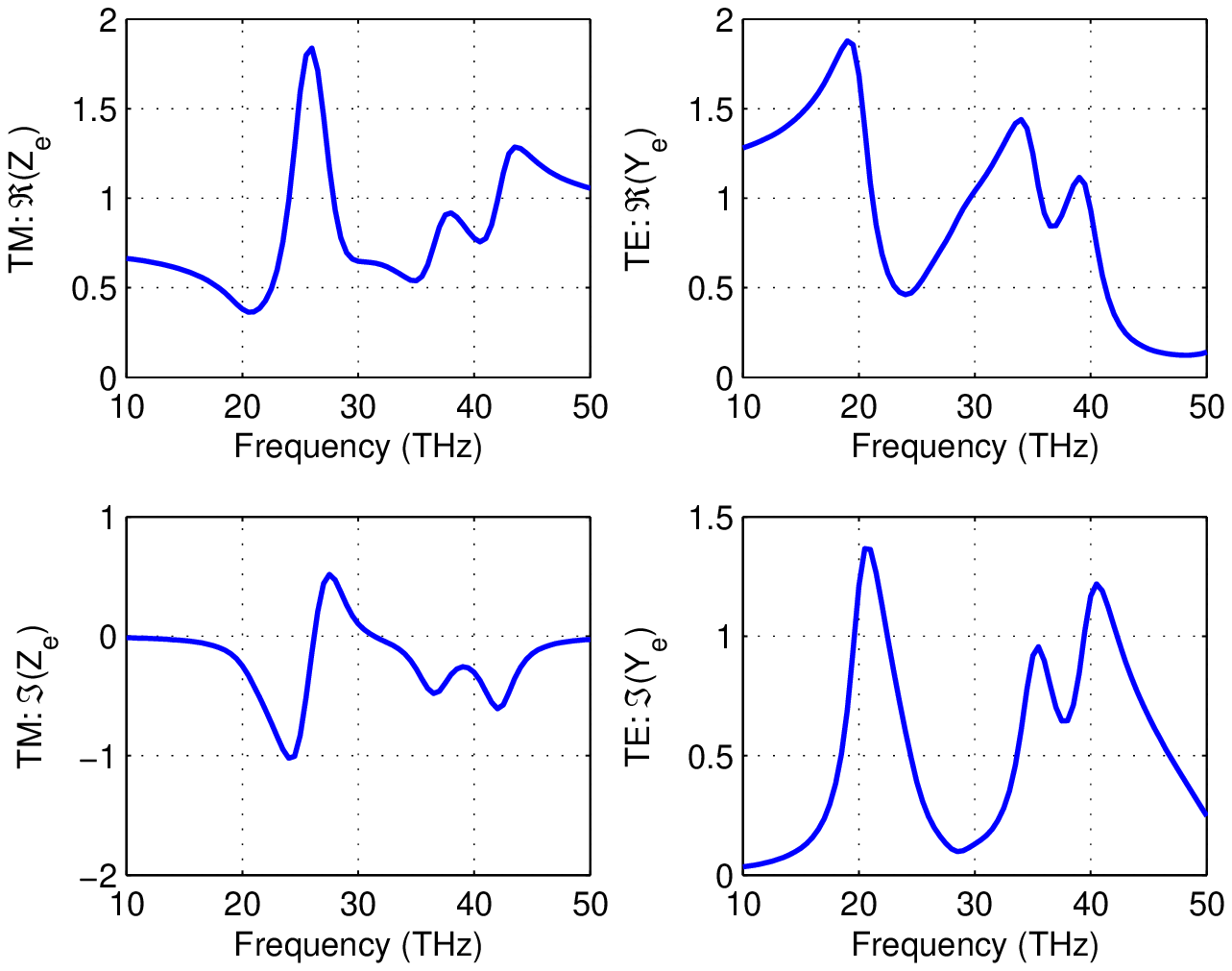}
}
\caption{Effective refractive index, impedance, and admittance retrieved from a single unit-cell layer.  Top panels: real part.  Bottom panels: imaginary part.}
\label{nZY}
\end{figure}

In this section, we provide an example to show how to implement the graphical method from the scattering parameters.  Typically the background is lossless and thus, the $X$ variable in Eq.~(\ref{RetvLine}) is real; whereas the slops and Y-intercepts are usually complex.  The real and imaginary parts of the retrieval lines should be plotted separately as shown in Fig.~\ref{RetrvLine} which was calculated from the scattering matrix.  Drude model is used for the effective material parameters of the equivalent layers A and B,
\begin{equation}
\label{Drude}
\epsilon = 1 - \frac{f_{ep}^2}{f^2-f_{er}^2+i\gamma f}  \,,\hspace{.3in}  \mu = 1 - \frac{f_{mp}^2}{f^2-f_{mr}^2+i\gamma f}  \,,
\end{equation}
where $f_{ep}=30\,$THz, $f_{mp}=20\,$THz, and $\gamma=3\,$THz.  The $\epsilon_x$ and $\mu_x$ are described by Eq.~(\ref{Drude}) with resonances at $f_{er}=20\,$THz and $f_{mr}=25\,$THz for the layer A and at $f_{er}=35\,$THz and $f_{mr}=37\,$THz for the layer B.  The other parameters for A: $\epsilon_y=\epsilon_x-0.3$, $\epsilon_z=\epsilon_x+2$, $\mu_y=\mu_x-0.5$, and $\mu_z=1$.  The other parameters for B: $\epsilon_y=\epsilon_x-0.8$, $\epsilon_z=\epsilon_x-0.5$, $\mu_y=\mu_x+0.2$, and $\mu_z=\mu_x-0.6$.  The thickness is $240\,$nm for the layer A and $320\,$nm for the layer B.  Thus, the period of the unit cell ($ABA$) is $800\,$nm.  Figure~\ref{Branch} shows the correct phase branch $m$ predicted for each frequency in the regime of interest for the thickness of one (a) and six (b) periods.  In our example since $d\ll\lambda$, for one unit-cell thickness, most of the frequencies are within the fundamental branch $m=0$ except for the regime of negative index of refraction [see Fig.~\ref{nZY}(a)] where $m=-1$.  For the six-period thickness, the phase branch jumps to $m=1$ in the frequency range $18\sim20\,$THz due to the high valves of the positive refractive index in this regime[see Fig.~\ref{nZY}(a)].  \\

\begin{figure}[hbt]
\centering
\subfigure[The effective $\overline\epsilon_x$ (left panels) and $\overline\mu_x$ (right panels).]
{
	\label{epmuxy:a}
	\includegraphics[width=.45\textwidth]{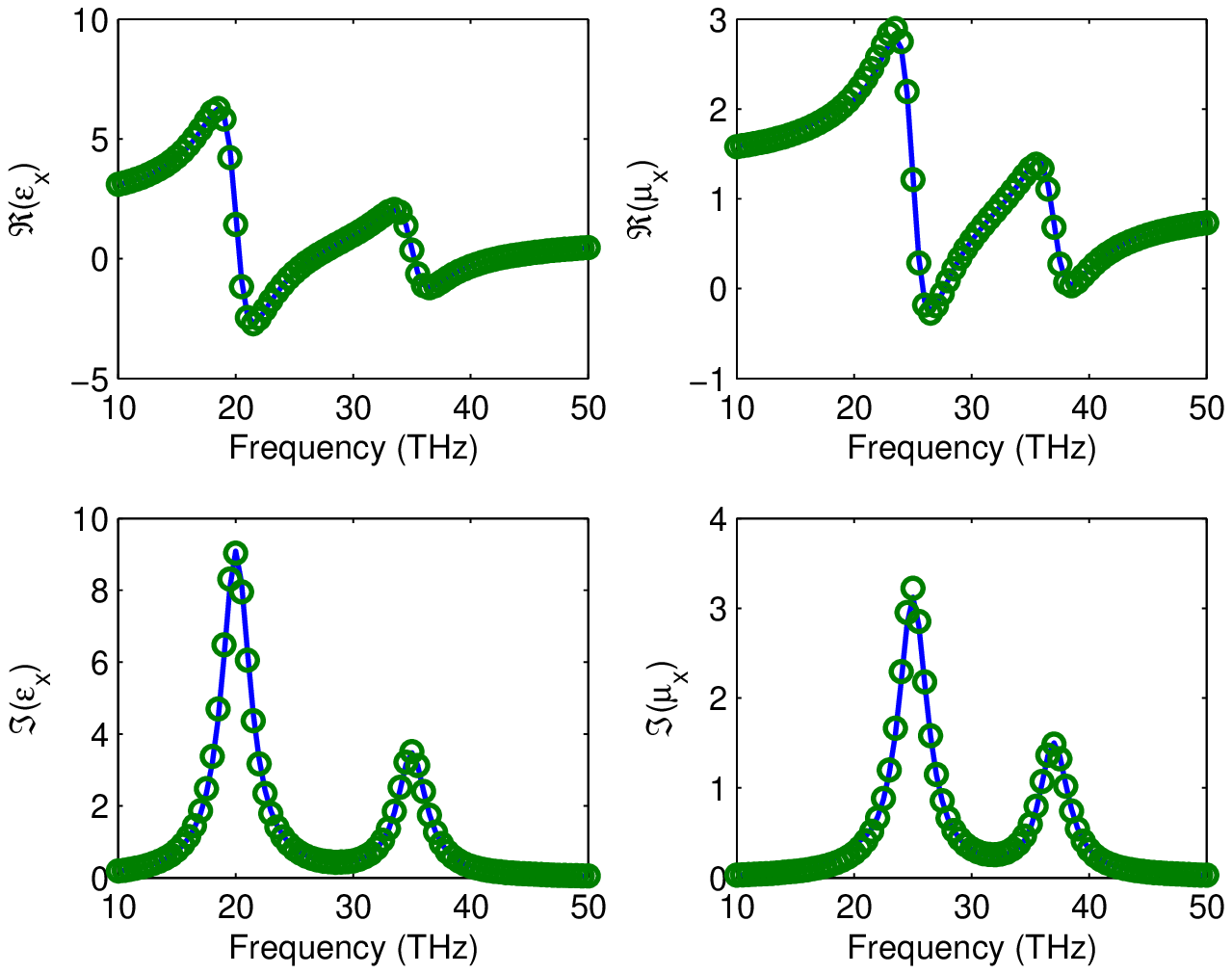}
}
\hspace{1cm}
\subfigure[The effective $\overline\epsilon_y$ (left panels) and $\overline\mu_y$ (right panels).]
{
	\label{epmuxy:b}
	\includegraphics[width=.45\textwidth]{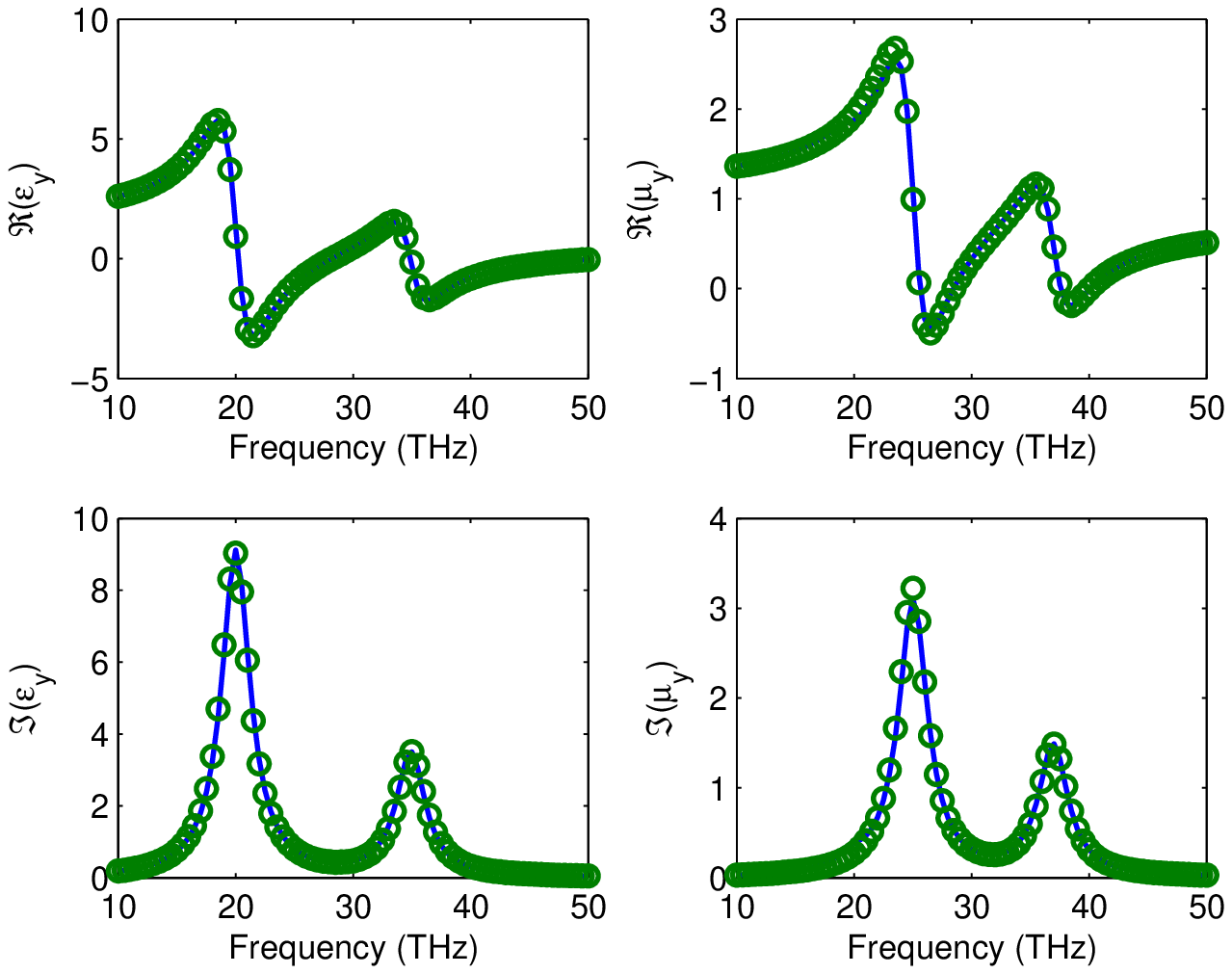}
}
\caption{In-plane values of the retrieved material parameters.  Upper panels: real parts.  Lower panels: imaginary parts.}
\label{epmuxy}
\end{figure}

\begin{figure}[h!]
\centering\includegraphics[width=.45\textwidth]{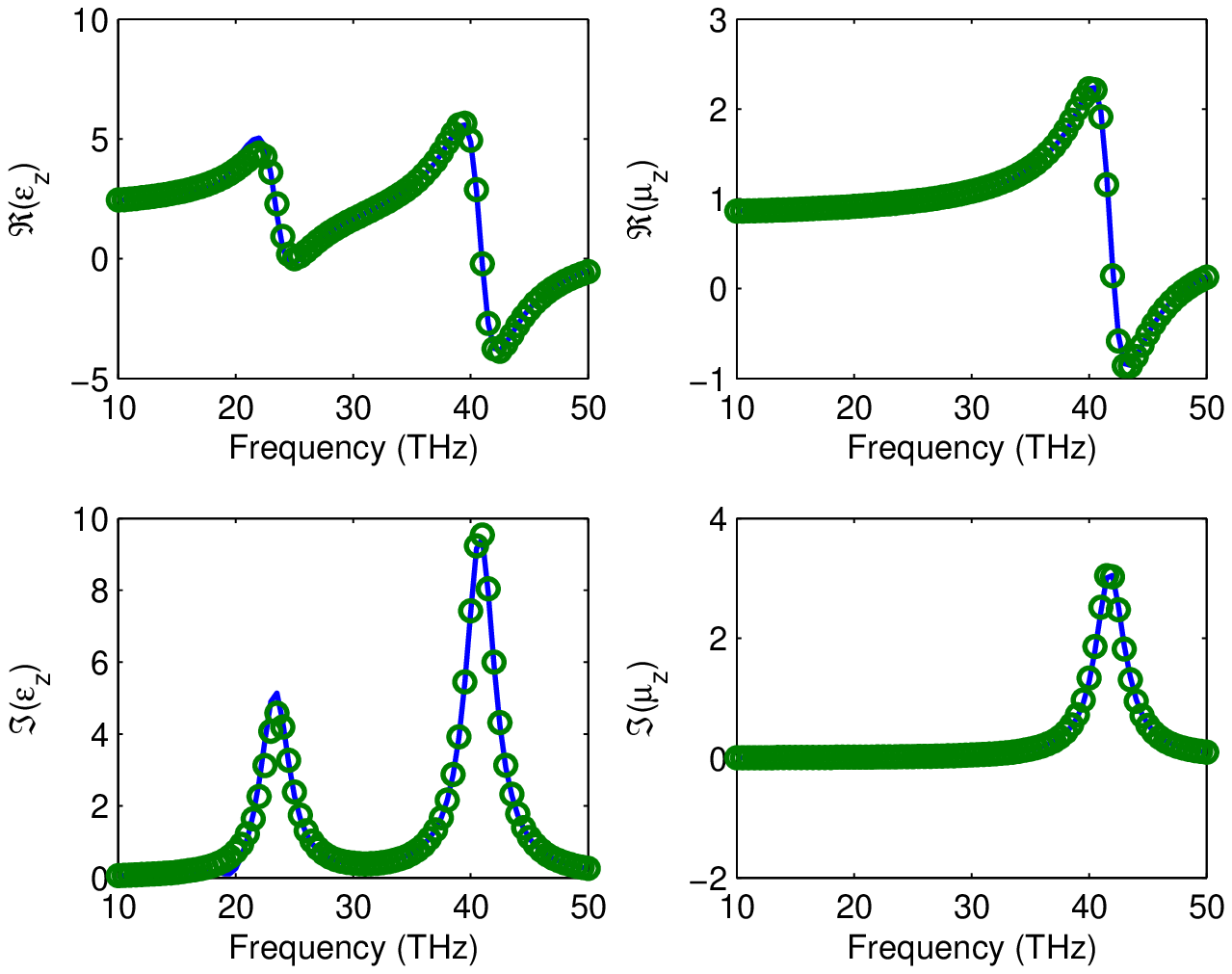}
\caption{The effective $\overline\epsilon_z$ (right panels) and $\overline\mu_z$ (left panels).  Upper panels: real parts.  Lower panels: imaginary parts.}
\label{epmuz}
\end{figure}

Shown in Fig.~\ref{nZY} are the effective index of refraction, impedance, and admittance retrieved from one unit cell.  These values are the same as those retrieved from the six unit cells.  In this example, both TM and TE polarizations contain two frequency bands where the effective index of refraction is negative.  As illustrated in Fig.~\ref{Branch}, these two negative bands are accurately captured by the branch-resolving techniques provided in Sec.~\ref{Phase}.  The six retrieved effective permittivities and permeabilities are shown in Figs.~\ref{epmuxy} and~\ref{epmuz} in the blue-solid curves, which are obtained from the graphical-retrieval and phase-unwrapping scheme, i.e. Eqs.~(\ref{RetVal}), (\ref{pmSign}), and~(\ref{Branch3}).  For comparison, we also show the effective permittivities and permeabilities calculated from Eq.~(\ref{epmuEff}) in Figs.~\ref{epmuxy} and~\ref{epmuz} in the green-circle curves.  The retrieved effective material parameters agree very well with the results calculated from the effective medium theory.  These values represent the bulk values of the material parameters since they are extracted from the symmetric unit cells.  We have successfully applied the graphical method and phase-unwrapping techniques to our recent experiment\cite{Roberts}.  From the experimental point of view, the straight-line graphical method is more accurate than the methods using single data point.  The linear regression technique is based on collective data points measured at several incidence angles.  It has an averaging effect and thus reduces the uncertainty of measurements.  Using several incidence angles can also help to resolve the phase ambiguity.  After determining the effective parameters of the bulk material, Eq.~(\ref{epmuEff}) can be used to recovery the effective parameters of the equivalent layers if we are interested in.  This extra benefit may help to improve the unit-cell design.  \\

When the tensors in Eq.~(\ref{epmu}) rotate about the z-axis without rotating the coordinate system, the off-diagonal elements in the x-y plane will not be zero ($\epsilon_{xy}\ne0$ and $\mu_{xy}\ne0$), and thus the x and y components of the electromagnetic fields will be coupled.  The dispersion relations in Eq.~(\ref{Disp}) will no longer be valid.  Without the proper modification, the current parameter retrieval scheme cannot be applied to this scenario.  The original classification of the left-handed and right-handed electromagnetic materials can cause confusion with chiral materials which are important class of electromagnetic materials\cite{McCall}.  To avoid such confusion, we can use phase-velocity characteristics.  When the phase velocity and the time-averaged Poynting vector form an acute angle, the medium is positive-phase-velocity characteristics.  When the phase velocity and the time-averaged Poynting vector form an obtuse angle, the medium is negative-phase-velocity characteristics.

\section{Conclusions}
In conclusions, we have shown that symmetric unit cells can often be constructed in metamaterials regardless the number of sub-layers and the complexity of each sub-layers in the original unit cells.  The graphical retrieval method and phase unwrapping techniques presented here might be a useful tool for metamaterial designs.

\section*{Acknowledgments}
The author thanks NAVAIR's ILIR program and the program managers Mark Spector and Steven Russell in Office of Naval Research for funding of this work.  \\


\begin{thebibliography}{99}

\bibitem{Fang}  N. Fang, H. Lee, C. Sun, and X. Zhang, ``Sub-diffraction-limited optical imaging with a silver superlens,'' Science {\bf308}, 534--537 (2005).

\bibitem{Valentine1}  J. Valentine, J. Li, T. Zentgraf, G. Bartal, and X. Zhang,  ``An optical cloak made of dielectrics,''  Nat. Mater. {\bf8}, 568--571 (2009).

\bibitem{Feng}  S. Feng and K. Halterman, ``Parametrically shielding electromagnetic fields by nonlinear metamaterials,'' \prl {\bf100}, 063901 (2008).

\bibitem{Capolino}  F. Capolino Ed. {\it Theory and phenomena of metamaterials} (CRC Press, Taylor and Francis Group, New York, 2009).


\bibitem{Valentine2}  J. Valentine, S. Zhang, T. Zentgraf, E. U. Avila, D. A. Genov, G. Bartal, and X. Zhang, ``Three-dimensional optical metamaterial with a negative refractive index,'' Nature (London) {\bf455}, 376--380 (2008).

\bibitem{NLiu}  N. Liu, H. Guo, L. Fu, S. Kaiser, H. Schweizer, and H. Giessen, ``Three-dimensional photonic metamaterials at optical frequencies,'' Nat. Mater. {\bf7}, 31--37 (2008).

\bibitem{Zhang}  S. Zhang, W. Fan, N. C. Panoiu, K. J. Malloy, R. M. Osgood, and S. R. J. Brueck,  ``Experimental demonstration of near-infrared negative-index metamaterials,''  \prl {\bf95}, 137404 (2005).

\bibitem{Shalaev}  V. M. Shalaev, W. Cai, U. K. Chettiar, H.-K. Yuan, A. K. Sarychev, V. P. Drachev, and A. V. Kildishev,  ``Negative index of refraction in optical metamaterials,''  \ol {\bf30}, 3356--3358 (2005).


\bibitem{Andryieuski}  A. Andryieuski, R. Malureanu, and A. V. Lavrinenko,  ``Wave propagation retrieval method for metamaterials: Unambiguous restoration of effective parameters,''  \prb {\bf80}, 193101 (2009).


\bibitem{Herpin}  A. Herpin, ``Calcul du pouvoir r\'{e}flecteur d'un syst\`{e}me stratifi\'{e} quelconque,''   Compt. Rend. {\bf225}, 182--183 (1947).


\bibitem{Smith}  D. R. Smith, S. Schultz, P. Marko$\check{\mbox{s}}$, and C. M. Soukoulis, ``Determination of effective permittivity and permeability of metamaterials from reflection and transmission coefficients,'' \prb {\bf65}, 195104 (2002).

\bibitem{Koschny}  T. Koschny, P. Marko$\check{\mbox{s}}$, E. N. Economou, D. R. Smith, D. C. Vier, and C. M. Soukoulis, ``Impact of inherent periodic structure on effective medium description of left-handed and related metamaterials,'' \prb {\bf71}, 245105 (2005).

\bibitem{Menzel}  C. Menzel, C. Rockstuhl, T. Paul, and F. Lederer, ``Retrieving effective parameters for metamaterials at oblique incidence,'' \prb {\bf77}, 195328 (2008).

\bibitem{Chen}  X. Chen, T. M. Grzegorczyk, B.-I. Wu, J. Pacheco, Jr., and J. A. Kong, ``Robust method to retrieve the constitutive effective parameters of metamaterials,'' \pre {\bf70}, 016608 (2004).


\bibitem{Epstein}  L. I. Epstein, ``The design of optical filters,'' \josa {\bf42}, 806--810 (1952).


\bibitem{Lakhtakia}  A. Lakhtakia,  ``Constraints on effective constitutive parameters of certain bianisotropic laminated composite materials,''  Electromagnetics {\bf29} 508--514 (2009).


\bibitem{Roberts} M. J. Roberts, S. Feng, M. Moran, and L. Johnson, ``Effective permittivity near zero in nanolaminates of silver and amorphous polycarbonate,''  J. Nanophoton. {\bf4}, 043511 (2010).

\bibitem{McCall}  M. W McCall, A. Lakhtakia, and W. S Weiglhofer, ``The negative index of refraction demystified,'' Eur. J. Phys. {\bf23}, 353--359 (2002).


\end{thebibliography}
\end{document}